\documentclass[a4paper]{article} 
\usepackage{graphicx, amssymb} 
\usepackage[sumlimits]{amsmath}
\usepackage{amsthm}
\usepackage{array}

\usepackage{fullpage}

\usepackage{hyperref}

\newtheorem{thm}{Theorem}

\newtheorem{lem}[thm]{Lemma}
\theoremstyle{remark}

\newtheorem{defn}[thm]{Definition}

\newcommand{\refsec}[1]{Section~\ref{sec:#1}}
\newcommand{\reffig}[1]{Figure~\ref{fig:#1}}
\newcommand{\refeqn}[1]{(\ref{eqn:#1})}
\newcommand{\reftbl}[1]{Table~\ref{tbl:#1}}
\newcommand{\refthm}[1]{Theorem~\ref{thm:#1}}
\newcommand{\reflem}[1]{Lemma~\ref{lem:#1}}
\newcommand{\refdefn}[1]{Definition~\ref{defn:#1}}
\newcommand{\reffoot}[1]{Footnote~\ref{foot:#1}}

\newcommand{\N}{\mathbb N}

\newcommand{\cG}{{\cal G}}
\newcommand{\cD}{{\cal D}}
\newcommand{\cB}{{\cal B}}
\newcommand{\cC}{{\cal C}}
\newcommand{\cZ}{{\cal Z}}

\newcommand{\ADV} {\mathrm{Adv}}

\newcommand{\s}[1]{\left(#1\right)}

\newcommand{\elem}[1]{[\![#1]\!]}

\newcommand{\pfstart}{\begin{proof}} 
\newcommand{\pfend}{\end{proof}} 

\newcommand{\stage}{s}

\title{Learning-Graph-Based Quantum Algorithm for $k$-distinctness}
\author{Aleksandrs Belovs\thanks{Faculty of Computing, University of Latvia, stiboh@gmail.com.}}
\date{}

\begin{document}
\maketitle

\begin{abstract}
We present a quantum algorithm solving the $k$-distinctness problem in $O\s{n^{1-2^{k-2}/(2^k-1)}}$ queries with a bounded error. This improves the previous $O(n^{k/(k+1)})$-query algorithm by Ambainis. The construction uses a modified learning graph approach. Compared to the recent paper by Belovs and Lee~\cite{lee:learningKdistPrior}, the algorithm doesn't require any prior information on the input, and the complexity analysis is much simpler.

Additionally, we introduce an $O(\sqrt{n}\alpha^{1/6})$ algorithm for the graph collision problem where $\alpha$ is the independence number of the graph.
\end{abstract}

\section{Introduction}
The {\em element distinctness} problem consists of computing function $f\colon [m]^n\to\{0,1\}$ that evaluates to 1 iff there is a pair of equal elements in the input, i.e., $f(x_1,\dots,x_n)=1$ iff $\exists i\ne j:  x_i=x_j$. (Here we use notation $[n]=\{1,2,\dots,n\}$.) The quantum query complexity of the element distinctness problem is well understood.  It is known to be $\Theta(n^{2/3})$, with the algorithm given by Ambainis~\cite{ambainis:distinctness}, and the lower bound shown by Aaronson and Shi~\cite{shi:collisionLower} and Kutin~\cite{kutin:collisionLower} for the case of large alphabet size $\Omega(n^2)$, and by Ambainis~\cite{ambainis:collisionLower} in the general case. 

Ambainis' algorithm for the element distinctness problem was the first application of the quantum random walk framework to a ``natural'' problem (i.e., one seemingly having little relation to random walks), and it had significantly changed the way quantum algorithms have been developed since then. The core of the algorithm is quantum walk on the Johnson graph. This primitive has been reused in many other algorithms: triangle detection in a graph given by its adjacency matrix~\cite{magniez:triangle}, matrix product verification~\cite{buhrman:productVerification}, restricted range associativity~\cite{dorn:associativity}, and others. Given that the behavior of quantum walk is well-understood for arbitrary graphs~\cite{szegedy:walk, magniez:walkSearch}, it is even surprising that the applications have been mostly limited to the Johnson graph.

The {\em $k$-distinctness problem} is a direct generalization of the element distinctness problem. Given the same input, the function evaluates to 1 iff there is a set of $k$ input elements that are all equal, i.e., a set of indices $a_1,\dots,a_k\in [n]$ with $a_i\ne a_j$ and $x_{a_i}=x_{a_j}$ for all $i\ne j$.

The situation with the quantum query complexity of the $k$-distinctness problem is not so clear. (In this paper we assume $k=O(1)$, and consider the complexity of $k$-distinctness as $n\to\infty$.) As element distinctness reduces to $k$-distinctness by repeating each element $k-1$ times, the lower bound of $\Omega(n^{2/3})$ carries over to the $k$-distinctness problem (this argument is attributed 
to Aaronson in Ref.~\cite{ambainis:distinctness}). This simple lower bound is the best known so far.

In the same paper~\cite{ambainis:distinctness} with the element distinctness algorithm, Ambainis applied quantum walk on the Johnson graph in order to solve the $k$-distinctness problem. This resulted in a quantum algorithm with query complexity  $O(n^{k/(k+1)})$. This was the best known algorithm for this problem prior to this paper. 

The aforementioned algorithms work by searching for a small subset of input variables such that the value of the function is completely determined by the values within the subset. For instance, the values of two input variables are sufficient to claim the value of the element distinctness function is 1, provided their values are equal. This is formalized by the notion of certificate complexity as follows.

An {\em assignment} for a function $f\colon \cD\to \{0,1\}$ with $\cD\subseteq [m]^n$ is a function $\alpha\colon S\to [m]$ with $S\subseteq [n]$. The {\em size} of $\alpha$ is $|S|$. An input $x=(x_i)\in [m]^n$ {\em satisfies} assignment $\alpha$ if $\alpha(i) = x_i$ for all $i\in S$. An assignment $\alpha$ is called a {\em $b$-certificate} for $f$, with $b\in\{0,1\}$, if $f(x)=b$ for any $x\in \cD$ satisfying $\alpha$. The {\em certificate complexity} $C_x(f)$ of $f$ on $x$ is defined as the minimal size of a certificate for $f$ that $x$ satisfies. The $b$-certificate complexity $C^{(b)}(f)$ is defined as $\max_{x\in f^{-1}(b)} C_x(f)$. Thus, for instance, 1-certificate complexity of element distinctness is 2, and 1-certificate complexity of triangle detection is 3.

Soon after the Ambainis' paper, it was realized~\cite{childs:subsetFinding} that the algorithm developed for $k$-distinctness can be used to evaluate, in the same number of queries, any function with 1-certificate complexity equal to $k$. Now we know that for some functions this algorithm is tight, due to the lower bound for the $k$-sum problem~\cite{spalek:kSumLower}. The goal of the $k$-sum problem is to detect, given $n$ elements of an Abelian group as input, whether there are $k$ of them that sum up to a prescribed element of the group. The $k$-sum problem is noticeable in the sense that, given any $(k-1)$-tuple of input elements, one has absolutely no information on whether they form a part of an (inclusion-wise minimal) 1-certificate, or not.

The aforementioned applications of the quantum walk on the Johnson graph (triangle finding, etc.) went beyond $O(n^{k/(k+1)})$ upper bound by utilizing additional relations between the input variables: the adjacency relation of the edges for the triangle problem, row-column relations for the matrix products, and so on. For instance, two edges in a graph can't be a part of a 1-certificate for the triangle problem, if they are not adjacent.

The $k$-distinctness problem is different in the sense that it doesn't possess any structure of the variables. But it does possess a relation between the {\em values} of the variables: two elements can't be a part of a 1-certificate if their values are different. However, it seems that quantum walk on the Johnson graph fails to utilize this structure efficiently.

In this paper, we use the learning graph approach to construct a quantum algorithm that solves the $k$-distinctness problem in $O\s{n^{1-2^{k-2}/(2^k-1)}}$ queries. Note that $O\s{n^{1-2^{k-2}/(2^k-1)}} = o(n^{3/4})$. Thus, our algorithm solves $k$-distinctness, for arbitrary $k$, in asymptotically less queries than the best previously known algorithm solves 3-distinctness.

The learning graph is a novel way of construction quantum query algorithms. Somehow, it may be thought as a way of designing a more flexible quantum walk than just on the Johnson graph. And compared to the quantum walk design paradigms from Ref.~\cite{szegedy:walk, magniez:walkSearch}, it is easier to deal with. In particular, it doesn't require any spectral analysis of the underlying graph. 

Up to date, the applications of learning graphs are as follows. Belovs~\cite{belovs:learning} introduced the framework and used it to improve the query complexity of triangle detection. Zhu~\cite{zhu:learning} and Lee, Magniez and Santha~\cite{lee:learningSubgraphs} extended this algorithm to the containment of arbitrary subgraphs. Belovs and Lee~\cite{lee:learningKdistPrior} developed an algorithm for the $k$-distinctness problem that beats the $O(n^{k/(k+1)})$-query algorithm given some prior information about the input. Belovs and Reichardt~\cite{belovs:learningClaws} use a construction resembling learning graph to obtain an optimal algorithm for finding paths and claws of arbitrary length in the input graph. Also, they deal with time-efficient implementation of learning graphs.

The paper is organized as follows. In \refsec{prelim} we define the (dual of the) adversary bound. It is the main technical tool underlying our algorithm. Also, we describe learning graphs and the previous algorithm for the $k$-distinctness problem.  In \refsec{outline}, we describe the intuition behind our algorithm, and describe the changes we have made to the model of the learning graph. In \refsec{collision} we give an algorithm for the graph collision problem as a preparation for the $k$-distinctness algorithm that we describe in Sections~\ref{sec:first} and~\ref{sec:final}.  Strictly speaking, Sections from~\ref{sec:model} to~\ref{sec:collision} are not necessary for understanding the $k$-distinctness algorithm: the proof in Sections~\ref{sec:first} and~\ref{sec:final} rely on \refthm{adversary} only.  However, these sections are necessary for understanding the intuition behind the algorithm.

\section{Preliminaries}
\label{sec:prelim}
In this paper, we are mainly concerned with query complexity of quantum algorithms, i.e., we measure the complexity by the number of queries to the input the algorithm makes in the worst case. For the definition of query complexity and its basic properties, a good reference is~\cite{buhrman:querySurvey}.

In \refsec{adversary} we describe a tight characterization of the query complexity by a relatively simple semi-definite program (SDP): the adversary bound, Eq.~\refeqn{adversary}. This is the main technical tool underlying our algorithm. 

Although Eq.~\refeqn{adversary} is an SDP, and thus can be solved in polynomial time in the size of the program, the latter is exponential in the number of variables, and becomes very hard to solve exactly as its size grows. The {\em learning graph}~\cite{belovs:learning} is a tool for designing feasible solutions to Eq.~\refeqn{adversary}, whose complexity is easier to analyze. We define it in Sections~\ref{sec:model} and~\ref{sec:goal}.  In the first one, we describe the model following Ref.~\cite{belovs:learning, lee:learningKdistPrior}.  In the second one, we describe a common way of constructing learning graphs for specific problems, and give an example of a learning graph for the $k$-distinctness problem corresponding to the Ambainis' algorithm.

\subsection{Dual adversary bound}
\label{sec:adversary}
The adversary bound, originally introduced by Ambainis \cite{ambainis:adversary}, is one of 
the most important lower bound techniques for quantum query complexity.  A strengthening of the adversary bound, known as the general adversary bound \cite{hoyer:adversaryNegative}, has recently been shown to characterize quantum query complexity, up to constant factors \cite{reichardt:adversaryTight, lee:stateConversion}.

The (general) adversary bound is a semi-definite program, and admits two equivalent formulations: the primal, used to prove lower bounds; and the dual, used in algorithm construction. We use the latter.

\begin{defn}
Let $f\colon  \cD \rightarrow \{0,1\}$ with $\cD\subseteq [m]^n$ be a function. The adversary bound $\ADV^\pm(f)$ is defined as the optimal value of the following optimization problem:
\begin{subequations}
\label{eqn:adversary}
\begin{alignat}{3}
&\mbox{\rm minimize} &\quad& \max_{x\in \cD}\sum\nolimits_{j \in [n]} X_j\elem{x,x} \label{eqn:advObjective}\\
& \mbox{\rm subject to}&& \sum\nolimits_{j\in[n]\colon x_j \ne y_j} X_j\elem{x,y} = 1 &\quad& \text{\rm whenever $ f(x) \ne f(y)$;} \label{eqn:advFeasible}\\
&&& X_j\succeq 0 && \mbox{\rm for all $j\in [n]$;}\label{eqn:advPositive}
\end{alignat}
\end{subequations}
where the optimization is over positive semi-definite matrices $X_j$ with rows and columns labeled by the elements of $\cD$, and $X\elem{x,y}$ is used to denote the element of matrix $X$ on the intersection of the row and column labeled by $x$ and $y$, respectively.
\end{defn}

The general adversary bound characterizes quantum query complexity. Let $Q(f)$ denote the query complexity of the best quantum algorithm evaluating $f$ with a bounded error.
\begin{thm}[\cite{hoyer:adversaryNegative, reichardt:adversaryTight, lee:stateConversion}]
\label{thm:adversary}
Let $f$ be as above.  Then, $Q(f)=\Theta(\ADV^\pm(f))$.
\end{thm}

\subsection{Learning graphs: Model-driven description}
\label{sec:model}
In this section we briefly introduce the simplest model of learning graph following Ref.~\cite{belovs:learning, lee:learningKdistPrior}.

\begin{defn}
\label{defn:learning}
A {\em learning graph} ${\cal G}$ on $n$ input variables is a directed acyclic connected graph with vertices labeled by subsets of $[n]$, the input indices. It has arcs connecting vertices labeled by $S$ and $S\cup\{j\}$ only, where 
$S\subseteq[n]$ and $j\in[n]\setminus S$. The root of ${\cal G}$ is the vertex labeled by $\emptyset$. Each arc $e$ is assigned positive real {\em weight} $w_e$.
\end{defn}

Note that it is allowed to have several (or none) vertices labeled by the same subset $S\subseteq[n]$. If there is unique vertex of $\cG$ labeled by $S$, we usually use $S$ to denote it. Otherwise, we denote the vertex by $(S,a)$ where $a$ is some additional parameter used to distinguish vertices labeled by the same subset $S$.

A learning graph can be thought of as a way of modeling the development of one's knowledge about the input during a query algorithm. Initially, nothing is known, and this is represented by the root labeled by $\emptyset$.  At a vertex labeled by $S\subseteq [n]$, the values of the variables in $S$ have been learned. Following an arc $e$ connecting vertices labeled by $S$ to $S\cup\{j\}$ can be interpreted as querying the value of variable $x_j$. We say the arc {\em loads} element $j$. When talking about a vertex labeled by $S$, we call $S$ the set of {\em loaded elements}.

The graph ${\cal G}$ itself has a very loose connection to the function being calculated. The following notion is the essence of the construction.

\begin{defn}
\label{defn:flow}
Let ${\cal G}$ be a learning graph on $n$ input variables, and $f:{\cal D}\to\{0,1\}$ be a function with domain ${\cal D}\subseteq [m]^n$. A {\em flow} on ${\cal G}$ is a real-valued function $p_e(x)$ where $e$ is an arc of ${\cal G}$ and $x\in f^{-1}(1)$. For a fixed input $x$, the flow $p_e = p_e(x)$ has to satisfy the following properties:
\begin{itemize}
\item vertex $\emptyset$ is the only source of the flow, and it has value 1. In other words, the sum of $p_e$ over all $e$ leaving $\emptyset$ is 1;
\item a vertex labeled by $S$ is a sink iff it contains a 1-certificate for $f$ on input $x$. Such vertices are called {\em accepting}. Thus, if $S\ne\emptyset$ and $S$ is not accepting then, for a vertex labeled by $S$, the sum of $p_e$ over all in-coming arcs equals the sum of $p_e$ over all out-going arcs.
\end{itemize}
\end{defn}

We always assume a learning graph ${\cal G}$ is equipped with a function $f$ and a flow $p$ that satisfy the constraints of \refdefn{flow}. Define the {\em negative complexity} of ${\cal G}$ and the {\em positive complexity for input $x\in f^{-1}(1)$} as
\begin{equation}
\label{eqn:learningComp1}
{\cal C}^0({\cal G}) = \sum_{e\in E} w_e\qquad\mbox{and}\qquad {\cal C}^1({\cal G}, x) = \sum_{e\in E} \frac{p_e(x)^2}{w_e},
\end{equation}
respectively, where $E$ is the set of arcs of ${\cal G}$. The {\em positive complexity} and the {\em (total) complexity} of ${\cal G}$ are defined as
\begin{equation}
\label{eqn:learningComp2}
{\cal C}^1({\cal G}) = \max_{x\in f^{-1}(1)} \cC^1(\cG, x)\qquad\mbox{and}\qquad {\cal C}({\cal G}) = \max\{{\cal C}^0({\cal G}), {\cal C}^1({\cal G})\},
\end{equation}
respectively.\footnote{Ref.~\cite{belovs:learning} defines $\cC(\cG)$ as $\sqrt{\cC^0(\cG)\cC^1(\cG)}$. Both definitions are equivalent, because one may make both $\cC^0(\cG)$ and $\cC^1(\cG)$ equal to $\sqrt{\cC^0(\cG)\cC^1(\cG)}$ by simultaneously scaling the weights of all the arcs by an appropriate coefficient.}  The following theorem links learning graphs and quantum query algorithms:

\begin{thm}[\cite{lee:learningKdistPrior}]
\label{thm:learning}
Assume $\cG$ is a learning graph for a function $f:{\cal D}\to \{0,1\}$ with ${\cal D}\subseteq [m]^n$. Then there exists a bounded-error quantum query algorithm for the same function with complexity $O(\cC(\cG))$.
\end{thm}

\pfstart[Proof sketch.] We reduce to \refthm{adversary}. For each arc $e$ from $S$ to $S\cup\{j\}$, we define a block-diagonal matrix $X^e_j = \sum_\alpha Y_\alpha$, where the sum is over all assignments $\alpha$ on $S$. Each $Y_\alpha$ is defined as $\psi\psi^*$ where, for each $z\in\cD$:
\[
\psi\elem{z} =
\begin{cases}
p_e(z)/\sqrt{w_e},& \mbox{$f(z)=1$, and $z$ satisfies $\alpha$;}\\
\sqrt{w_e},& \text{$f(z)=0$, and $z$ satisfies $\alpha$;}\\
0,&\mbox{otherwise.}
\end{cases}
\]
Finally, we define $X_j$ in~\refeqn{adversary} as $\sum_e X_j^e$ where the sum is over all arcs $e$ loading $j$.

Condition~\refeqn{advPositive} is trivial, and the expression for the objective value~\refeqn{advObjective} is straightforward to check. The feasibility~\refeqn{advFeasible} is as follows. Fix any $x\in f^{-1}(1)$ and $y\in f^{-1}(0)$. By construction, $X_j^e\elem{x,y}=p_e(x)$, if $x_S = y_S$ where $S$ is the origin of $e$; otherwise, it is zero. Thus, only arcs $e$ from $S$ to $S\cup\{j\}$, such that $x_S=y_S$ and $x_j\ne y_j$, contribute to the sum in~\refeqn{advFeasible}. These arcs define a cut between the source $\emptyset$ and all the sinks of the flow $p_e=p_e(x)$, hence, the total value of the flow on these arcs is 1, as required.
\pfend

\subsection{Learning graphs: Procedure-driven description}
\label{sec:goal}
In this section, we describe a way of designing learning graphs that was used in  Ref.~\cite{belovs:learning} and other papers.  The learning graph, introduced in \refsec{model}, may be considered as a randomized procedure for loading values of the variables with the goal of convincing someone the value of the function is 1.  For each input $x\in f^{-1}(1)$, the designer of the learning graph builds its own procedure.  The goal is to load a 1-certificate for $x$.  Usually, for each positive input, one specific 1-certificate is chosen.  The elements inside the certificate are called {\em marked}.  The procedure is not allowed to err, i.e., it always has to load all the marked elements in the end.  The value of the complexity of the learning graph arises from the interplay between the procedures for different inputs. 


We illustrate this concepts with an example of a learning graph corresponding to the $k$-distinctness algorithm by Ambainis~\cite{ambainis:distinctness}.  Fix a positive input $x$, i.e., one evaluating to 1. Let $M=\{a_1,a_2\dots,a_k\}$ be such that $x_{a_1}=x_{a_2}=\cdots=x_{a_k}$. It is a 1-certificate for $x$.  The elements inside $M$ are marked.  One possible way of loading the marked elements consists of $k+1$ stage and is given in \reftbl{old}.  The internal randomness of the procedure is concealed in the choice of the $r$ elements on stage I.  (Here $r=o(n)$ is some parameter to be specified later.)  Each choice has probability $q={n-k\choose r}^{-1}$.

\begin{table}[htb]
\centering
\begin{tabular}{rp{13cm}}
\hline
I.& Load $r$ elements different from $a_1,\dots,a_k$.\\
II.1& Load $a_1$.\\
II.2& Load $a_2$.\\
& \quad\vdots\\
II.$k$& Load $a_k$.\\
\hline
\end{tabular}
\caption{Learning graph for the $k$-distinctness problem corresponding to the algorithm from Ref.~\cite{ambainis:distinctness}.}
\label{tbl:old}
\end{table}

Let us describe how a graph $\cG$ and flow $p$ is constructed from the description in \reftbl{old}. At first, we define the {\em key vertices} of $\cG$. If $d$ is the number of stages, the key vertices are $V_0\cup\cdots\cup V_d$, where $V_0=\{\emptyset\}$ and $V_i$ consists of all possible sets of variables loaded after $i$ stages.

For a fixed input $x$ and fixed internal randomness, the sets $S_{i-1}\in V_{i-1}$ and $S_i\in V_i$ of variables loaded before and after stage $i$, respectively, are uniquely defined. In this case, we connect $S_{i-1}$ and $S_i$ by a {\em transition} $e$.%
\footnote{In Ref.~\cite{belovs:learning}, the graph formed by the key vertices and the transitions is called reduced learning graph.}
 For that, we choose an arbitrary order $t_1,\dots,t_\ell$ of elements in $S_i\setminus S_{i-1}$, and connect $S_{i-1}$ and $S_i$ by a path:
\[
S_{i-1}, (S_{i-1}\cup\{t_1\},e), (S_{i-1}\cup\{t_1,t_2\},e),\dots, (S_i\setminus\{t_{\ell}\},e), S_i
\]
in $\cG$. Here, additional labels $e$ in the internal vertices assure that the paths corresponding to the transitions do not intersect, except at the ends. We say transition $e$ and all arcs therein belong to stage $i$.

\begin{figure}[tbh] 
\centering 
\includegraphics[width=8cm]{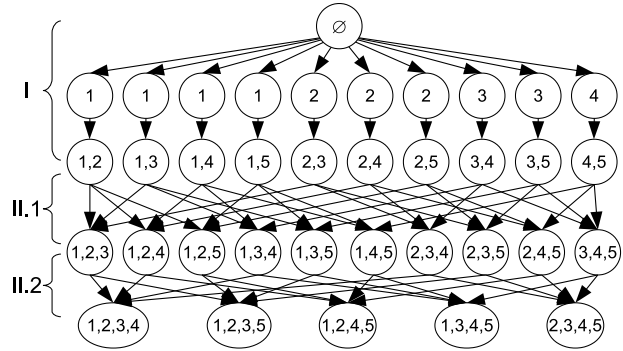} 
\caption{The learning graph for $k$-distinctness from \reftbl{old} in the case $k=2$, $n=5$ and $r=2$. Stages I, II.1 and II.2 shown.}
\label{fig:distinct}
\end{figure}

In the case like in the previous paragraph, we say the transition $e$ is {\em taken} for this choice of $x$ and the randomness. We say a transition is {\em used} for input $x$, if it is taken for some choice of the internal randomness. The set of transitions of $\cG$ is the union of all transitions used for all inputs in $f^{-1}(1)$. For instance, stage II.2 of the learning graph from \reftbl{old} consists of all transitions from $S$ to $S\cup\{j\}$ where $|S|=r+1$ and $j\notin S$.  For an example refer to \reffig{distinct}.

The flow $p_e(x)$ is defined as the probability, over the internal randomness, that transition $e$ is taken for input $x$. All arcs forming the transition are assigned the same flow. Thus, the transition $e$ is used by $x$ iff $p_e(x)>0$. In the learning graph from \reftbl{old}, $p_e(x)$ attains two values only: 0 and $q$.

So far, we have constructed the graph $\cG$ and the flow $p$. It remains to define the weights $w_e$. This is done using \refthm{symmetric} below. But, for that, we need some additional notions.

The {\em length} of stage $i$ is the number of variables loaded on this stage, i.e., $|S_{i}\setminus S_{i-1}|$ for a transition $e$ from $S_{i-1}$ to $S_i$ of stage $i$. In our applications in this paper this number is independent on the choice of $e$. We say the flow is {\em symmetric} on stage $i$ if the non-zero value of $p_e(x)$ is the same for all $e$ on stage $i$ and all $x$.%
\footnote{This is a less general definition than in Ref.~\cite{belovs:learning}, but it suffices for our purposes.}
 The flow in the learning graph from \reftbl{old} is symmetric.

If the flow is symmetric on stage $i$, we define the {\em speciality} $T_i$ of stage $i$ as the ratio of the total number of transitions on stage $i$, to the number of ones used by $x$.   In a symmetric flow, this quantity doesn't depend on $x$.

Finally, we define the {\em (total) complexity of stage $i$}, $\cC_i(\cG)$, similarly as $\cC(\cG)$ is defined in~\refeqn{learningComp1} and~\refeqn{learningComp2} with the summation over $E_i$, the set of all arcs on stage $i$, instead of $E$. It is easy to see that $\cC(\cG)$ is at most $\sum_i \cC_i(\cG)$.

\begin{thm}[\cite{belovs:learning}]
\label{thm:symmetric}
If the flow is symmetric on stage $i$, the arcs on stage $i$ can be weighted so that the complexity of the stage becomes $L_i\sqrt{T_i}$.
\end{thm}

\pfstart[Proof sketch]
Let $q$ be the non-zero value of the flow on stage $i$. Assign weight $q/\sqrt{T_i}$ to all arcs on stage $i$.
\pfend

Now we are able to calculate the complexity of the learning graph in \reftbl{old}. The length of stage I is $r$, and the length of stage II.$i$ is 1 for all $i$. It is also not hard to see that the corresponding specialities are $O(1)$ and $O(n^{i}/r^{i-1})$.  For example, a transition from $S$ to $S\cup\{j\}$ on stage II.$k$ is used by input $x$ iff $a_1,\dots,a_{k-1}\in S$ and $j=a_k$. For a random choice of $S$ and $j\notin S$, the probability of $j=a_k$ is $1/n$, and the probability of $a_1,\dots,a_{k-1}\in S$, given $j=a_{k}$, is $\Omega(r^{k-1}/n^{k-1})$. Thus, the total probability is $\Omega(r^{k-1}/n^k)$ and the speciality is the inverse of that.

Thus, the complexity of the algorithm, by Theorems~\ref{thm:symmetric} and~\ref{thm:learning}, is $O(r+\sqrt{n^k/r^{k-1}})$. It is optimized when $r = n^{k/(k+1)}$, and the complexity is $O(n^{k/(k+1)})$.

\section{Outline of the algorithm}
\label{sec:outline}
In this section we describe how the learning graph from \reftbl{old} is transformed into a new learning graph with a better complexity. Many times when learning graphs were applied to new problems, they were modified accordingly~\cite{belovs:learning, lee:learningKdistPrior, belovs:learningClaws}. This paper is not an exception, thus, we also describe the modifications we make to the model of a learning graph.

The main point of the learning graph in \reftbl{old} and similar ones is to reduce the speciality of the last step, loading $a_k$. In the learning graph from \reftbl{old}, it is achieved by loading $r$ non-marked elements before loading the certificate. This way, the speciality of the last step gets reduced from $O(n^k)$ to $O(n^k/r^{k-1})$. We say that $a_1,\dots,a_{k-1}$ are {\em hidden} among the $r$ elements loaded on stage I. The larger the set we hide the elements into, the better.

Unfortunately, we can't make $r$ as large as we like, because loading the non-marked elements also counts towards the complexity. At the equilibrium point $r=n^{k/(k+1)}$, we attain the optimal complexity of the learning graph.

In Ref.~\cite{lee:learningKdistPrior} a learning graph was constructed with better complexity. It uses a more general version of the learning graph than in \refsec{model}, with weights of the arcs dependent on the values of the element loaded so far.  Its main idea is to hide $a_1,\dots,a_{k-1}$ as one entity, not $k-1$ independent elements. By gradually distilling vertices of the learning graph having large number of $(k-1)$-tuples of equal elements, the learning graph manages to reduce the speciality of the last step without increasing the number of elements loaded, because $\{a_1,\dots,a_{k-1}\}$ gets hidden among a relatively large number of $(k-1)$-tuples of equal elements.

But this learning graph has serious drawbacks. Due to dealing with the values of the variables in the distilling phase, the flow through the learning graph ceases to be symmetric and depends heavily on the input. This makes the analysis of the learning graph quite complicated. What is even worse, the learning graph requires strong prior knowledge on the structure of the input to attain reasonable complexity.

In this paper we construct a learning graph that combines the best features of both learning graphs. Its complexity is the same as in Ref.~\cite{lee:learningKdistPrior}. Also, it has the flow symmetric and almost independent on the input, like the one in \reftbl{old}. This has three advantages compared to the learning graph in Ref.~\cite{lee:learningKdistPrior}: its complexity is easier to analyze, it doesn't require any prior information on the input, and it is more suitable for a time-efficient implementation along the lines of Ref.~\cite{belovs:learningClaws}. This is achieved at the cost of a more involved construction.

Let us outline the modifications the learning graph from \reftbl{old} undergoes in order to reduce the complexity. Again, we assume $x$ is a positive input, and $M=\{a_1,\dots,a_k\}$ is such that $x_{a_1}=\cdots=x_{a_k}$.

\begin{enumerate}
\item\label{intro} We achieve a symmetric flow with smaller speciality of the last step by finding a way to load more non-marked elements in the first stages of the learning graph.  There is an indication that it is possible in some cases: the values of $r$ Boolean variables can be learned in less than $r$ queries, if there is a bias between the number of ones and zeros~\cite{boyer:groverTight}. More precisely, if the number of ones is $\ell$, the values can be loaded in $O(\sqrt{r\ell})$ queries.

\item\label{division} We start with dividing the set $S$ of loaded elements into $k$ subsets: $S=S_1\sqcup \cdots\sqcup S_{k-1}$, where $\sqcup$ denotes disjoint union. Set $S_i$ has size $r_i=o(n)$. We use $S_i$ to hide $a_i$ when loading $a_k$. This step doesn't reduce the speciality, but this division will be necessary further.

\item\label{star} Consider the situation before loading $a_k$. If an element $j\in S_2$ is such that $x_j\ne x_t$ for all $t\in S_1$, this element cannot be a part of the certificate (i.e., it can't be $a_2$), and its precise value is irrelevant. (This is the place where we utilize the relations between the values of the variables as mentioned in the introduction.) In this case, we say $j$ doesn't have a {\em match} in $S_1$, and represent it by a special symbol $\star$.  Otherwise, we {\em uncover} the element, i.e., load its precise value.  Similarly, when loading $S_i$ with $i>2$, we uncover those elements only that have a match among the uncovered elements of $S_{i-1}$.


\item\label{bias} Usually, the number of elements in $S_i$ having a match in $S_{i-1}$ is much smaller than the total number of elements in $S_i$. Similarly to Point~\ref{intro}, we can reduce the complexity of loading elements in $S_i$ because of this bias. 
Thus, we have $r_i = \omega(r_{1})$, while the complexity of loading remains $O(r_1)$. Now we have more elements to hide $a_i$ in between, hence, the speciality of loading $a_k$ gets reduced.

\item\label{outline} When loading $a_k$, we do want $a_i$ to be in $S_i$ for $i\in[k-1]$, because that is where we hide them. On the other hand, in order to keep the speciality of loading non-marked elements in $S_1,\dots,S_{k-1}$ equal to $O(1)$, we would like to add $a_1$ to $S_1$ only after all elements in $S_{k-1}$ have been already loaded. Thus, we load $a_1,\dots,a_{k-1}$ between these two stages and put them in $S_1,\dots,S_{k-1}$. This is summarized in \reftbl{new}.

\item\label{problem} Since the uncovering of elements in $S_i$, for $i>1$, depends on the values contained in $S_j$ with $j<i$, adding $a_i$ to $S_i$ afterwards is a bit of cheating. This does cause some problems we describe in more detail in \refsec{firstFeasible}. We describe a solution in \refsec{final}.


\end{enumerate}

\begin{table}
\begin{tabular}{rp{13cm}}
\hline
I.1& Load a set $S_1$ of $r_1$ elements not from $M$.\\
I.2& Load a set $S_2$ of $r_2$ elements not from $M$, uncovering only those elements that have a match in $S_1$.\\
I.3& Load a set $S_3$ of $r_3$ elements not from $M$, uncovering only those elements that have a match among the uncovered elements of $S_2$.\\
& \quad\vdots\\
I.($k-1$)& Load a set $S_{k-1}$ of $r_{k-1}$ elements not from $M$, uncovering only those elements that have a match among uncovered elements of $S_{k-2}$.\\
II.1& Load $a_1$ and add it to $S_1$.\\
& \quad\vdots\\
II.$(k-1)$& Load $a_{k-1}$ and add it to $S_{k-1}$.\\
II.$k$& Load $a_k$.\\
\hline\end{tabular}
\caption{An illustrative (not correct) version of the learning graph for $k$-distinctness}
\label{tbl:new}
\end{table}

In order to account for these changes, we use the following modifications to the learning graph model.
\renewcommand{\theenumi}{\Alph{enumi}}
\begin{enumerate}
\item\label{drop} In \refsec{final}, we are forced to drop the flow notion from \refdefn{flow}.%
\footnote{
The reader should not be confused by our earlier statement that the flow is symmetric, because when considering one stage, the part of the ``flow'' is still symmetric.  It only is not defined where it comes from, and where it goes afterwards. See also \reffoot{symmetric}. 
}  
We use \refthm{adversary} directly, borrowing some concepts from the proof of \refthm{learning}.  Namely, the notion of a vertex and an arc leaving it.  Also, we keep the internal randomness intuition from \refsec{goal}.  The loading procedure still doesn't err in some sense formalized in~\refeqn{sum}.

\item\label{vertices} We change the way the vertices of the learning graph are represented. Firstly, we keep track to which $S_i$ each loaded element belongs, like said in Point~\ref{division}. Also, we assume the condition on uncovering of elements, and use the special symbol $\star$ as a notation for a covered element, as described in Point~\ref{star}.  Technically, this corresponds to modification of the definition of an assignment $\alpha$ in $Y_\alpha$ in the proof of \refthm{learning}.

\item\label{rank} Instead of having a rank-1 matrix $Y_\alpha$ as in the proof of \refthm{learning}, we define it as a rank-2 matrix.  The weight of the arc depends now on the value of the variable being loaded as well, although in a rather restricted form.  Thus, we are able to make use of the bias as described in Point~\ref{bias}, and to account for the introduction of $\star$ in Point~\ref{star}.
\end{enumerate}
\renewcommand{\theenumi}{\arabic{enumi}}

The remaining part of the paper is organized as follows.  In \refsec{collision} we give a learning graph for the graph collision problem that uses some ideas from above (Points~\ref{intro} and~\ref{rank}).  In Sections~\ref{sec:first} and~\ref{sec:final} we describe the algorithm for $k$-distinctness.  In order to simplify the exposition, we first give a version of the learning graph from \reftbl{new} that illustrates the main idea of the algorithm, but has a flaw.  We identify it in \refsec{firstFeasible} and then describe a work-around in \refsec{final}.  The complexity analysis of the second algorithm is analogous to the first one, so we do it for the first algorithm.

\section{Warm-up: Graph collision}
\label{sec:collision}
In order to get ready for the $k$-distinctness algorithm, we start with a learning graph for the graph collision problem with an additional promise. It is a learning graph version of the algorithm by Ambainis~\cite{ambainis:personal}. 

The graph collision problem is one of the ingredients of the triangle finding quantum algorithm by Magniez {\em et al.}~\cite{magniez:triangle} and the learning-graph-based quantum algorithm by Belovs~\cite{belovs:learning}. It is also used in the algorithm for boolean matrix multiplication by Jeffery {\em et al.}~\cite{jeffery:matrixMultiplication}.

The problem is parametrized by a simple graph $G$ on $n$ vertices. The input is formed by $n$ boolean variables: one for each vertex of the graph. The function evaluates to 1 if there exists an edge of $G$ with both endpoints marked by value 1, and to 0 otherwise.

The best known quantum algorithm solving this problem for a general graph $G$ uses $O(n^{2/3})$ queries. For specific classes of graphs one can do better. For instance, if $G$ is the complete graph, graph collision is equivalent to the 2-threshold problem that can be solved in $O(\sqrt{n})$ queries by two applications of the Grover algorithm. The algorithm in this section may be interpreted as an interpolation between this trivial special case and the general case.

Recall that the independence number $\alpha(G)$ of a simple graph $G$ is the maximal cardinality of a subset of vertices of $G$ such that no two of them are connected by an edge.

\begin{thm}
\label{thm:collision}
Graph collision on graph $G$ can be solved in $O(\sqrt{n}\alpha^{1/6})$ quantum queries with bounded error, where $\alpha=\alpha(G)$ is the independence number of $G$.
\end{thm}

Note that if $G$ is a complete graph, $\alpha(G)=1$, and we get the previously mentioned $O(\sqrt{n})$-algorithm for this trivial case. In the general case, $\alpha(G)=O(n)$, and the complexity of the algorithm is $O(n^{2/3})$ that coincides with the complexity of the algorithm for a general graph.

Jeffery {\em et al.}~\cite{jeffery:matrixMultiplication} build a quantum algorithm solving graph collision on $G$ in $O(\sqrt{n}+\sqrt{m})$ queries if $G$ misses $m$ edges to be a complete graph. This algorithm is incomparable to the one in \refthm{collision}: for some graphs the algorithm from \refthm{collision} performs better, for some graphs, vice versa.

\pfstart[Proof of \refthm{collision}] 
Let $f$ be the graph collision function specified by graph $G$. The first step of the algorithm is quantum counting~\cite{brassard:counting}. We distinguish the case when the number of ones in the input is at most $\alpha$, and when it is at least $2\alpha$. In the intermediate case, the counting subroutine is allowed to return any of the outcomes. The complexity of the subroutine is $O(\sqrt{n})$. 

If we know, with high probability, that the number of ones is greater than $\alpha$, we may claim that graph collision exists. Otherwise, we may assume the number of ones is at most $2\alpha$. In this case, we execute the following learning graph $\cG$.

The learning graph is essentially the learning graph from \reftbl{old} for 2-distinctness. Let us denote, for simplicity, $a=a_1$ and $b=a_2$. Then, instead of loading $a$ and $b$ such that $x_a=x_b$, the graph collision learning graph loads $a$ and $b$ such that $x_a=x_b=1$ and $ab$ is an edge of $G$. We reduce the complexity of the learning graph by utilizing the bias between the number of zeros and ones induced by the small independence number, as outlined in Points~\ref{intro} and~\ref{rank} of \refsec{outline}. 

One could prove the correctness of the algorithm completely analogously to the correctness proof of the algorithm from \refsec{goal}. However, in the preparation for future discard of the notion of flow (Point~\ref{drop} from \refsec{outline}), we use language from \refsec{final}. The reader is encouraged to compare both ways of the proof.

Let $x$ be a positive input, and let $a$ and $b$ be such that $x_a=x_b=1$ and $ab$ is an edge of $G$. Set $M=\{a,b\}$ is a 1-certificate for $x$.

The key vertices of the learning graph are $V_1\cup V_2$, where $V_1$ and $V_2$ consist of all subsets of $[n]$ of sizes $r$ and $r+1$, respectively, where $r=o(n)$ is some parameter to be specified later.%
\footnote{
Compared to the learning graph for 2-distinctness from \refsec{goal}, we do not have $V_0$ and $V_3$. The reason for the absence of $V_0$ is described in \reffoot{merge}. Set $V_3$ is omitted because no arc originates there, hence, by the view of Point~\ref{drop} from \refsec{outline}, it is of no importance for us.}

A vertex in $V_1$ completely specifies the internal randomness. For each $R\in V_1$, we fix an arbitrary order of its elements: $R=\{t_1,\dots, t_r\}$. We say the choice of randomness $R\in V_1$ is {\em consistent} with $x$ if $\{a,b\}\cap R=\emptyset$. For each $x\in f^{-1}(1)$, there are exactly ${n-2\choose r}$ choices of $R\in V_1$ consistent with $x$. We take each of them with probability $q={n-2\choose r}^{-1}$. 

For a fixed input $x$ and fixed randomness $R=\{t_1,\dots,t_r\}\in V_1$ consistent with $x$, the elements are loaded (we are going to define what this means later) in the following order: 
\begin{equation}
\label{eqn:collisionT}
t_1,\dots,t_r,t_{r+1}=a,t_{r+2} = b.
\end{equation}

The non-key vertices of $\cG$ are of the form $v=(\{t_1,\dots,t_\ell\}, R)$, where $0\le\ell<r$, $R\in V_1$, and $t_i$ are from~\refeqn{collisionT}. Recall that, as stated in \refsec{model}, the first element of the pair is the set of loaded elements, and the second one is an additional mark used to distinguish vertices with the same set of loaded elements.

An {\em arc} of the learning graph is a process of loading one variable. We denote it by $A^v_j$. Here, $j$ is the variable the arc loads, and $v$ is a vertex of $\cG$ it originates in. In our case, the arcs are as follows. The arcs of the stage I have $v=(\{t_1,\dots,t_\ell\}, R)$ and $j=t_{\ell+1}$ with $0\le \ell<r$.%
\footnote{
\label{foot:merge} These arcs may be considered as in transitions from $\emptyset$ to the elements of $V_1$. In order to obtain a learning graph similar to the one in \reffig{distinct}, one has to merge all vertices of the form $(\emptyset, R)$ into one vertex $\emptyset$ forming $V_0$.}
 The arcs of stages II.1 and II.2 have $v=S$, with $S\in V_1$ and $S\in V_2$, respectively, and $j\notin S$.

For a fixed $x\in f^{-1}(1)$ and fixed internal randomness $R\in V_1$ consistent with $x$, the arcs taken are
\begin{equation}
\label{eqn:collisionTaken}
A^{(\{t_1,\dots,t_\ell\},R)}_{t_{\ell+1}}\; \mbox{for $0\le \ell<r$,}\qquad A^R_a\qquad\text{and}\qquad A^{R\cup\{a\}}_b.
\end{equation}
Recall, we say $x$ {\em satisfies} an arc if the arc is taken for some $R\in V_1$ consistent with $x$. Note also, no arc is taken for two different choices of the randomness.

Like in the proof of \refthm{learning}, for each arc $A^v_j$, we assign a matrix $X^v_j\succeq 0$. Then, $X_j$ in \refeqn{adversary} are given by $X_j = \sum_v X^v_j$. 

Fix $A^v_j$, and let $S$ be the set of loaded elements. Recall that an assignment on $S$ as a function $\alpha\colon S\to \{0,1\}$.  An input $z\in \{0,1\}^n$ {\em satisfies} assignment $\alpha$ iff $z_t=\alpha(t)$ for each $t\in S$. We say inputs $x$ and $y$ {\em agree} on $S$, if they satisfy the same assignment $\alpha$. Let $X^v_j = \sum_\alpha Y_\alpha$ where the sum is over all assignments $\alpha$ on $S$. The matrix $Y_\alpha$ is defined as $q(\psi\psi^*+\phi\phi^*)$, where, for each $z\in\{0,1\}^n$,
\[
\psi\elem{z}=
\begin{cases}
1/\sqrt{w_1},& \parbox{4.3cm}{$f(z)=1$, $z_j=1$,\\$z$ satisfies $\alpha$ and the arc $A^v_j$;}\\[\bigskipamount]
\sqrt{w_1},& \parbox{4cm}{$f(z)=0$, $z_j=0$,\\and $z$ satisfies $\alpha$;}\\[\medskipamount]
0,&\mbox{otherwise;}
\end{cases}
\mbox{and}\qquad
\phi\elem{z}=
\begin{cases}
1/\sqrt{w_0},& \parbox{4.3cm}{$f(z)=1$, $z_j=0$,\\$z$ satisfies $\alpha$ and the arc $A^v_j$;}\\[\bigskipamount]
\sqrt{w_0},& \parbox{4cm}{$f(z)=0$, $z_j=1$,\\ and $z$ satisfies $\alpha$;}\\[\medskipamount]
0,&\mbox{otherwise.}
\end{cases}
\]
Here $w_0$ and $w_1$ are parameters to be specified later (the weights of the arc). They depend only on the stage the arc belongs to. In other words, $X^v_j$ consists of the blocks of the following form:
\begin{equation}
\label{eqn:blockCollision}
\begin{array}{r|cc|cc|}
& x_j=1 & x_j=0 & y_j=1 & y_j=0 \\
\hline
x_j=1 & q/w_1 & 0 & 0 & q \\
x_j=0 & 0 & q/w_0 & q & 0\\
\hline
y_j=1 & 0 & q & qw_0 & 0  \\
y_j=0 & q & 0 & 0 & qw_1\\
\hline
\end{array}
\end{equation}
Here each of the 16 elements corresponds to a block in $Y_\alpha$ with all entries equal to this element. The first and the second columns represent the elements from $f^{-1}(1)$ that satisfy $\alpha$ and $A^v_j$, and such that their $j$th element equals $1$ and $0$, respectively. Similarly, the third and the fourth columns represent elements from $f^{-1}(0)$ that satisfy $\alpha$ and such that their $j$th element equals $1$ and $0$, respectively. This construction is due to Robin Kothari~\cite{kothari:personal}.

\subsection{Feasibility} 
Assume $x$ and $y$ are inputs such that $f(x)=1$ and $f(y)=0$. Let $R\in V_1$ be a choice of the internal randomness consistent with $x$. Let $Z_j$ be the matrix corresponding to the arc loading $j$ that is taken for the input $x$ and randomness $R$. I.e., 
$Z_j$ is either the matrix of~\refeqn{collisionTaken} with sub-index $j$, or $Z_j=0$, if there are none, i.e., when $j\notin R\cup\{a,b\}$. We are going to prove that
\begin{equation}
\label{eqn:collisionSum}
\sum_{j: x_j\ne y_j} Z_j\elem{x,y} = q.
\end{equation}
(This is what we meant by saying in Point~\ref{drop} of \refsec{outline} that the learning graph doesn't err for all choices of the internal randomness.) Since there are ${n-2\choose r}$ choices of $R$ consistent with $x$, and no arc is taken for two different choices of the randomness, this proves the feasibility condition in~\refeqn{advFeasible}.

Consider the order~\refeqn{collisionT} in which elements are loaded for this particular choice of $x$ and $R$. Before any element is loaded, both inputs agree (they satisfy the same assignment $\alpha\colon\emptyset\to\{0,1\}$). After all elements are loaded, $x$ and $y$ disagree, because it is not possible that $y_a=x_a$ and $y_b=x_b$. With each element loaded, the assignments become more specific. This means that there exists an element $j=t_i$ such that $x$ and $y$ agree before loading $j$, but disagree afterwards. In particular, $x_j\ne y_j$. By construction, this $j$ contributes $q$ to the sum in~\refeqn{collisionSum}.  All other $j$ contribute 0 to the sum. Indeed, if $j'=t_{i'}$ with $i'<i$ then $x_{j'}=y_{j'}$, hence, $j'$ contributes 0.  For $j'=t_{i'}$ with $i'>i$, $x$ and $y$ disagree on $\{t_1,\dots,t_{i'-1}\}$, hence, $Z_{j'}\elem{x,y}=0$ by construction.

\subsection{Complexity} 
\label{sec:collisionComplexity}
Similarly to \refsec{goal}, let us define the complexity of stage $i$ on input $z\in\{0,1\}^n$ as $\sum_{j\in [n]} X'_j\elem{z,z}$, where $X'_j = \sum_v X^v_j$ with the sum over $v$ such that $A^v_j$ belongs to stage $i$. Also, define the complexity of stage $i$ as the maximum complexity over all inputs $z\in\{0,1\}^n$. Clearly, the objective value~\refeqn{advObjective} of the whole program is at most the sum of the complexities of all stages.

Let us start with stages II.1 and II.2.%
  \footnote{For stages II.1 and II.2, the complexity of the stage can be calculated using \refthm{symmetric} like in \refsec{goal}. For stage II.1, the length is 1, and the speciality is $O(n)$. For stage II.2, the length is 1, and the speciality is $O(n^2/r)$. Hence, the complexities are $O(\sqrt{n})$ and $O(n/\sqrt{r})$, respectively.}
  For any $x\in f^{-1}(1)$, on both stages II.1 and II.2 there are ${n-2\choose r}$ arcs satisfying it.  These are arcs $A^R_a$ and $A^{R\cup\{a\}}_b$, respectively, for all choices of $R\in V_1$ consistent with $x$.  By~\refeqn{blockCollision}, each of them contributes $q/w_1$ to the complexity of $x$ on stages II.1 and II.2, respectively.  Since, we are guaranteed that $x_j=1$ in notations from~\refeqn{blockCollision}, we may set $w_0=0$. 

The total number of arcs on stages II.1 and II.2 are $(n-r){n\choose r}$ and $(n-r-1){n\choose r+1}$, respectively. Each of them contributes at most $qw_1$ to the complexity of any $y\in f^{-1}(0)$ on stages II.1 and II.2, respectively.

Thus, the complexities of stages II.1 and II.2 on any $x\in f^{-1}(1)$ is ${n-2\choose r}q/w_1=1/w_1$. On any $y\in f^{-1}(0)$, it is at most $(n-r){n\choose r}qw_1 = O(nw_1)$ and $(n-r-1){n\choose r+1}qw_1 = O(n^2w_1/r)$, respectively. If we set $w_1$ equal to $1/\sqrt{n}$ on stage II.1 and to $\sqrt{r}/n$ on stage II.2, the complexities of these stages become $O(\sqrt{n})$ and $O(n/\sqrt{r})$, respectively.

Consider stage I now. Let $k$ be the number of variables with value 1 in the input ($x$ or $y$). The total number of arcs on this stage is $r{n\choose r}$. Out of them, exactly $k{n-1\choose r-1}$ load a variable with value 1. Thus, for $y\in f^{-1}(0)$, the complexity of stage I is
\[qr{n\choose r}w_0 + qk{n-1\choose r-1}w_1 = O\s{rw_0 + \frac{kr}{n}w_1}. \]
Similarly, for $x\in f^{-1}(1)$, the complexity of stage I is $O(r/w_1+kr/(nw_0))$. If we set $w_0=\sqrt{\alpha/n}$ and $w_1 =\sqrt{n/\alpha}$ then, since $k\le 2\alpha$, the complexity of stage I becomes $O(r\sqrt{k/n})$.  The total complexity of the learning graph is
\[O\s{r\sqrt{\frac{\alpha}{n}} + \frac{n}{\sqrt{r}}} = O\s{\sqrt{n}\alpha^{1/6}},\]
if $r=n\alpha^{-1/3}$. 
\pfend

\section{Algorithm for $k$-distinctness: First attempt}
\label{sec:first}
The aim of this and the next sections is to prove the following theorem:
\begin{thm}
\label{thm:main}
For arbitrary but fixed integer $k\ge2$, the $k$-distinctness problem can be solved by a quantum computer in $O\s{n^{1-2^{k-2}/(2^k-1)}}$ queries with a bounded error.
\end{thm}

As mentioned in \refsec{outline}, we do not rely on previous results like \refthm{symmetric} in the proof, and use \refthm{adversary} directly.  The construction of the algorithm deviates from the graph representation: a bit in \refsec{first}, and quite strongly in \refsec{final}.  However, we keep the term ``vertex'' for an entity describing some knowledge of the values of the input variables, and the term ``arc'' for a process of loading a value of a variable (possibly, only partially). Each arc originates in a vertex, but we do not specify where it goes.  Inspired by \refsec{goal}, the vertices are divided into {\em key} ones denoted by the {\em set of loaded variables} $S$ with additional structure.  The non-key vertices are denoted by $(S,R)$ where $S$ is the set of loaded variables, and $R$ is an additional label used to distinguish vertices with the same $S$, as described in \refsec{model}.  Also, we use the ``internal randomness'' term from \refsec{goal}.

Throughout Sections~\ref{sec:first} and~\ref{sec:final}, let $f:[m]^n\to\{0,1\}$ be the $k$-distinctness function.  The section is organized as follows.  In \refsec{firstConstruction}, we rigorously define the learning graph from \reftbl{new}; in \refsec{complexity}, analyze its complexity; and, finally, describe the flaw mentioned in Point~\ref{problem} of \refsec{outline} in \refsec{firstFeasible}.

Similarly to the analysis in Ref.~\cite{ambainis:distinctness}, we may assume there is unique $k$-tuple of equal elements in any positive input.%
\footnote{\label{foot:overkill} Actually, this is an overkill: as we will see from the proof, it is enough for our algorithm to assume there are at most $O(n)$ pairs of equal elements in the input, that is a weaker assumption.}
One of the simplest reductions to this special case is to take a sequence $T_i$ of uniformly random subsets of $[n]$ of sizes $(2k/(2k+1))^in$, and to run the algorithm, for each $i$, with the input variables outside $T_i$ removed. One can prove that if there are $k$ equal elements in the input then there exists $i$ such that, with probability at least $1/2$, $T_i$ will contain unique $k$-tuple of equal elements. The complexities of the executions of the algorithm for various $i$ form a geometric series, and their sum is equal to the complexity of the algorithm for $i=0$ up to a constant factor. Refer to Ref.~\cite{ambainis:distinctness} for more detail and alternative reductions.

\subsection{Construction}
\label{sec:firstConstruction}
The construction of the learning graph $\cG$ for $k$-distinctness is similar to the one in \refthm{collision}. Let $x$ be a positive input, and let $M=\{a_1,a_2,\dots,a_k\}$ denote the unique $k$-tuple of equal elements in $x$. The key vertices of the learning graph are $V_1\cup\cdots \cup V_k$, where $V_\stage$, for $\stage\in[k]$, consists of all $(k-1)$-tuples $S=(S_1,\dots,S_{k-1})$ of pairwise disjoint subsets of $[n]$ of the following sizes.  For $V_\stage$, we require that $|S_i|=r_i+1$ for $i<\stage$, and $|S_i|=r_i$ for $i\ge \stage$.

Again, a vertex $R=(R_1,\dots,R_{k-1})\in V_1$ completely specifies the internal randomness.  We assume that, for any $R\in V_1$, an arbitrary order $t_1,\dots, t_r$ of the elements in $\bigcup R = R_1\cup \cdots \cup R_{k-1}$ is fixed so that all elements of $R_i$ precede all elements of $R_{i+1}$ for all $i\le k-2$. (Here $r=\sum_i r_i$.) We say $R\in V_1$ is {\em consistent} with $x$ if $\{a_1,\dots,a_k\}\cap (\bigcup R)=\emptyset$. 

For each $x\in f^{-1}(1)$, there are exactly ${n-k\choose r_1,\dots,r_{k-1}}$ choices of $R\in V_1$ consistent with $x$.  We take each of them, in the sense of \refsec{goal}, with probability $q={n-k\choose r_1,\dots,r_{k-1}}^{-1}$. Here we use notation
\[
{N\choose b_1,\dots,b_i} = {N\choose b_1}{N-b_1\choose b_2}\cdots{N-b_1-\cdots-b_{i-1}\choose b_i}.
\]

For a fixed input $x$ and fixed randomness $R\in V_1$ consistent with $x$, the elements are loaded in the following order: 
\begin{equation}
\label{eqn:t}
t_1,t_2,\dots,t_r,t_{r+1}=a_1,t_{r+2} = a_2,\dots,t_{r+k}=a_k.
\end{equation}

We use a similar convention to name the vertices and the arcs of the learning graph as in \refthm{collision}. The non-key vertices of $\cG$ are of the form $v=(R\cap \{t_1,\dots,t_\ell\}, R)$, where $R\in V_1$, $0\le \ell<r$, and $\{t_i\}$ are from~\refeqn{t}. Here we use notation $R\cap T = (R_1\cap T,\dots,R_{k-1}\cap T)$. The first element of the pair describes the set of loaded elements.

Let us describe the arcs $A^v_j$ of $\cG$, where, again, $j$ is the variable the arc loads, and $v$ is the vertex of $\cG$ it originates in. The arcs of the stages I.$\stage$ have $v=(R\cap \{t_1,\dots,t_\ell\}, R)$ and $j=t_{\ell+1}$ with $0\le \ell<r$. The arc belongs to stage I.$\stage$ iff $t_{\ell+1}\in R_\stage$. The arcs of stage II.$\stage$ have $v=S$, with $S\in V_\stage$, and $j\notin \bigcup S$.

For a fixed $x\in f^{-1}(1)$ and fixed internal randomness $R\in V_1$ consistent with $x$, the following arcs are {\em taken}: 
\begin{equation}
\label{eqn:taken}
A^{(R\cap \{t_1,\dots,t_\ell\}, R)}_{t_{\ell+1}}\; \mbox{for $0\le \ell<r$}\qquad\text{and}\qquad A^{R[a_1,\dots,a_\ell]}_{a_{\ell+1}}\; \text{with $0\le\ell < k$}.
\end{equation}
Here
\[R[a_1,a_2,\dots,a_\ell] = (R_1\cup\{a_1\},R_2\cup\{a_2\}\dots,R_\ell\cup\{a_\ell\},R_{\ell+1},\dots,R_{k-1}).\] 
We say $x$ {\em satisfies} all these arcs. Note that, for a fixed $x$, no arc is taken for two different choices of $R$.

Again, for each arc $A^v_j$, we assign a matrix $X^v_j\succeq 0$, so that $X_j$ in \refeqn{adversary} are given by $X_j = \sum_v X^v_j$. Assume $A_j^v$ is fixed. Let $S=(S_1\dots,S_{k-1})$ be the set of loaded elements. Define an assignment on $S$ as a function $\alpha\colon \bigcup S \to [m]\cup\{\star\}$, where $\star$ represents the {\em covered} elements of stages I.$\stage$ for $\stage>1$. Thus, $\alpha$ must satisfy $\star\notin \alpha(S_1)$ and $\alpha(S_{i+1}) \subseteq \alpha(S_i)\cup\{\star\}$ for $1\le i\le k-2$. An input $z\in [m]^n$ {\em satisfies} assignment $\alpha$ iff, for each $t\in \bigcup S$,
\[\alpha(t) = \begin{cases}
z_t,& \mbox{$t\in S_1$; or $t\in S_i$ for $i>1$ and $z_t\in \alpha(S_{i-1})$;}\\
\star,& \mbox{otherwise}.
\end{cases}\]
Each input $z$ satisfies unique assignment on $S$. Again, we say inputs $x$ and $y$ {\em agree} on $S$, if they satisfy the same assignment on $S$.

We define $X^v_j$ as $\sum_\alpha Y_\alpha$ where the sum is over all assignments $\alpha$ on $S$. The definition of $Y_\alpha$ depends on whether $A^v_j$ is on stage I.$\stage$ with $\stage>1$, or not. If $A^v_j$ is not on one of these stages then $Y_\alpha = q \psi \psi^*$ where, for each $z\in [m]^n$,
\[
\psi\elem{z} = \begin{cases}
1/{\sqrt{w}},& \text{$f(z)=1$, and $z$ satisfies $\alpha$ and the arc $A_j^v$;}\\
\sqrt{w},& \text{$f(z)=0$, and $z$ satisfies $\alpha$;}\\
0,&\text{otherwise.}
\end{cases}
\]
Here $w$ is a positive real number: the weight of the arc. It only depends on the stage of the arc, and will be specified later. Thus, $X_j^v$ consists of the blocks of the following form:
\begin{equation}
\label{eqn:tablica1}
\begin{array}{r|cc|}
& x & y \\
\hline
x & q/w & q\\
y & q & qw\\
\hline
\end{array}
\end{equation}
Here $x$ and $y$ represent inputs mapping to 1 and 0, respectively, all satisfying some assignment $\alpha$. The inputs represented by $x$ have to satisfy the arc $A_j^v$ as well. 

If $A_j^v$ is on stage I.$\stage$ with $\stage>1$, the elements having a match in $S_{\stage-1}$ and the ones that don't must be treated differently. In this case, $Y_\alpha=q(\psi\psi^*+\phi\phi^*)$, where
\[
\psi\elem{z}=
\begin{cases}
1/\sqrt{w_1},& \parbox{4.5cm}{$f(z)=1$, $z_j\in \alpha(S_{\stage-1})$,\\and $z$ satisfies $\alpha$ and $A_j^v$;}\\[\medskipamount]
\sqrt{w_1},& \mbox{$f(z)=0$, and $z$ satisfies $\alpha$};\\
0,&\mbox{otherwise};
\end{cases}
\qquad
\phi\elem{z}=\begin{cases}
1/\sqrt{w_0},& \parbox{5cm}{$f(z)=1$, $z_j\notin \alpha(S_{\stage-1})$,\\and $z$ satisfies $\alpha$ and $A_j^v$;}\\[\bigskipamount]
\sqrt{w_0},& \parbox{5cm}{$f(z)=0$, $z_j\in \alpha(S_{\stage-1})$,\\ and $z$ satisfies $\alpha$;}\\[\medskipamount]
0,&\mbox{otherwise}.
\end{cases}
\]
Here $w_0$ and $w_1$ are again parameters to be specified later. In other words, $X_j^v$ consists of the blocks of the following form:
\begin{equation}
\label{eqn:tablica}
\begin{array}{r|cc|cc|}
& x_j\in \alpha(S_{\stage-1}) & x_j\notin \alpha(S_{\stage-1}) & y_j\in \alpha(S_{\stage-1}) & y_j\notin\alpha(S_{\stage-1}) \\
\hline
x_j\in \alpha(S_{\stage-1}) & q/w_1 & 0 & q & q \\
x_j\notin \alpha(S_{\stage-1}) & 0 & q/w_0 & q & 0\\
\hline
y_j\in \alpha(S_{\stage-1}) & q & q & q(w_0+w_1) & qw_1  \\
y_j\notin \alpha(S_{\stage-1}) & q & 0 & qw_1 & qw_1\\
\hline
\end{array}
\end{equation}
Here $x$ and $y$ are like in~\refeqn{tablica1}. This is a generalization of the construction from \refthm{collision}. Note that if $x_j$ and $y_j$ are both represented by $\star$ in the assignments on $(S_1,\dots,S_{s-1},S_s\cup\{j\},S_{s+1},\dots,S_{k-1})$ they satisfy then $X^v_j\elem{x,y}=0$.

\subsection{Complexity}
\label{sec:complexity}
Let us estimate the complexity of the learning graph. We use the notion of the complexity of a stage from \refsec{collisionComplexity}.

Let us start with stage I.1. We set $w=1$ for all arcs on this stage. There are $r_1{n\choose r_1,\dots,r_{k-1}}$ arcs on this stage, and, by~\refeqn{tablica1}, each of them contributes at most $q$ to the complexity of each $z\in\{0,1\}^n$. Hence, the complexity of stage I.1 is $O\s{qr_1{n\choose r_1,\dots,r_{k-1}}} = O(r_1)$.

Now consider stage II.$\stage$ for $\stage\in[k]$.%
\footnote{\label{foot:symmetric} The complexities of stages I.1 and II.$\stage$ can be explained by a similar argument like in \refsec{goal}. For stage I.1, the length is $r_1$, and the speciality is $O(1)$. For stage II.$\stage$, the length is 1, but the speciality is $O(n^\stage/(r_1\cdots r_{\stage-1}))$, because there are $\stage$ marked elements involved, giving $O(n^\stage)$, but $a_i$, for $i<\stage$, is hidden in $S_i$ of size $r_i$, hence, the speciality gets divided by $r_1\cdots r_{\stage-1}$. This argument works, because the ``flow'' is symmetric (the $(x,y)$-entries of $X^v_j$ are either 0 or $q$) as highlighted in \refsec{outline}. }
   The total number of arcs on the stage is $(n-r-\stage+1){n\choose r_1+1,\dots,r_{\stage-1}+1,r_\stage,\dots,r_{k-1}}$.  By~\refeqn{tablica1}, each of them contribute $qw$ to the complexity of each $y\in f^{-1}(0)$.  Out of these arcs, for any $x\in f^{-1}(1)$, exactly ${n-k\choose r_1,\dots,r_{k-1}}$ satisfy $x$.  And each of them contribute $q/w$ to the complexity of $x$.  Thus, the complexities of stage II.$\stage$ for any input in $f^{-1}(0)$ and $f^{-1}(1)$ are
\[(n-r-\stage+1){n\choose r_1+1,\dots,r_{\stage-1}+1,r_\stage,\dots,r_{k-1}}qw = O\s{\frac{n^\stage w}{r_1\cdots r_{\stage-1}}}, \]
and
\[{n-k\choose r_1,\dots,r_{k-1}}\frac{q}{w} = \frac{1}{w}, \]
respectively.  By setting $w = \s{n^s/(r_1\cdots r_{\stage-1})}^{-1/2}$, we get complexity $O\s{\sqrt{n^s/(r_1\cdots r_{\stage-1})}}$ of stage II.$\stage$. The maximal complexity is attained for stage II.$k$.

Now let us calculate the complexity of stage I.$\stage$ for $\stage>1$.  The total number of arcs on this stage is $r_\stage{n\choose r_1,\dots,r_{k-1}}$.  Consider an input $z\in [m]^n$, and a choice of the internal randomness $R=(R_1,\dots,R_{k-1})\in V_1$.  An element $j$ is uncovered on stage I.$\stage$ for this choice of $R$ if and only if there is an $\stage$-tuple $(b_1,\dots,b_\stage)$ of elements with $j=b_\stage$ such that $b_i\in R_i$ and $z_{b_i}=z_{b_j}$ for all $i, j\in[\stage]$.  By our assumption on the uniqueness of a $k$-tuple of equal elements in a positive input, the total number of such $\stage$-tuples is $O(n)$.  And, for each of them, there are ${n-\stage \choose r_1-1,\dots,r_\stage-1,r_{\stage+1},\dots,r_{k-1}}$ choices of $R\in V_1$ such that $b_i\in R_i$ for all $i\in[s]$.  By~\refeqn{tablica}, the complexities of this stage for an input in $f^{-1}(0)$ and in $f^{-1}(1)$ are, respectively, at most
\[q\left[O(n) {n-\stage\choose r_1-1,\dots,r_\stage-1,r_{\stage+1},\dots,r_{k-1}}w_0+r_\stage{n\choose r_1,\dots,r_{k-1}}w_1 \right]=
O\s{\frac{r_1\cdots r_\stage}{n^{\stage-1}}w_0+r_\stage w_1} \]
and
\[q\left[O(n) {n-\stage\choose r_1-1,\dots,r_\stage-1,r_{\stage+1},\dots,r_{k-1}}\frac{1}{w_1}+r_\stage{n\choose r_1,\dots,r_{k-1}}\frac1{w_0} \right]=
O\s{\frac{r_1\cdots r_\stage}{n^{\stage-1}w_1}+\frac{r_\stage}{w_0}}. \]
By assigning $w_0=\sqrt{n^{\stage-1}/(r_1\cdots r_{\stage-1})}$ and $w_1 = \sqrt{r_1\cdots r_{\stage-1}/n^{\stage-1}}$, both these quantities become $O\s{r_\stage\sqrt{r_1\cdots r_{\stage-1}/n^{\stage-1}}}$.  

With this choice of the weights, the value of the objective function in~\refeqn{advObjective} is
\begin{equation}
\label{eqn:complexity}
O\s{r_1+ r_2\sqrt{\frac{r_1}{n}}+\cdots+r_{k-1}\sqrt{\frac{r_1\cdots r_{k-2}}{n^{k-2}}} + \sqrt{\frac{n^k}{r_1\cdots r_{k-1}}}}.
\end{equation}
Assuming all terms in~\refeqn{complexity} except the last one are equal, and denoting $\rho_i = \log_n r_i$, we get that
\[\rho_i +\frac12(\rho_1+\cdots+\rho_{i-1}) - \frac{i-1}{2} = \rho_{i+1} + \frac12(\rho_1+\cdots+\rho_{i}) - \frac{i}{2},\qquad\mbox{for $i=1,\dots,k-2$;} \]
or, equivalently,
\[\rho_{i+1} = \frac{1 + \rho_{i}}2,\qquad\mbox{for $i=1,\dots,k-2$.} \]
Assuming the first term, $r_1$, equals the last one, $\sqrt{n\frac{n}{r_1}\cdots\frac{n}{r_{k-1}}}$, we get
\[ \rho_1 = \frac{1+(1-\rho_1)+\cdots+(1-\rho_{k-1})}2 = \frac12 + \s{\frac12 + \cdots + \frac{1}{2^{k-1}}}(1-\rho_1) =  \frac12 + \s{1-\frac{1}{2^{k-1}}}(1-\rho_1).\]
From here, it is straightforward that $\rho_1 = 1 - 2^{k-2}/(2^k-1)$, hence, the complexity of the algorithm is $O\s{n^{1-2^{k-2}/(2^k-1)}}$.

\subsection{(In)feasibility}
\label{sec:firstFeasible}
 Assume $x$ and $y$ are inputs such that $f(x)=1$ and $f(y)=0$. Let $R=(R_1,\dots,R_{k-1})\in V_1$ be a choice of the internal randomness consistent with $x$. Similarly to the proof of \refthm{collision}, let $Z_j$ be the matrix corresponding to the arc loading $j$ that is taken for input $x$ and randomness $R$ (i.e., the one from~\refeqn{taken} with sub-index $j$, or the zero matrix, if there are none). 

Again, we would like to prove that~\refeqn{collisionSum} holds. Unfortunately, it doesn't always hold. Assume $x$, $y$ and $R\in V_1$ are such that $x$ and $y$ agree on $R$. Thus, the contribution to~\refeqn{collisionSum} is 0 from all arcs of stages I.$\stage$. Now assume that $x_{a_1}=y_{a_1}$ and there exists $b\in R_2$ such that $y_b = x_{a_1}$. This doesn't contradict that $x$ and $y$ agree on $R$, because $y_b$ is represented by $\star$ in the assignment it satisfies on $R$. 

But $x$ and $y$ disagree on $R[a_1]$, because $y_b$ gets uncovered there. Thus, the contribution to~\refeqn{collisionSum} is 0 from all arcs of stages II.$\stage$ as well. Thus, equation~\refeqn{collisionSum} doesn't hold. We deal with this problem in the next section.

\section{Final version}
\label{sec:final}
In \refsec{firstFeasible}, we saw that the learning graph in \reftbl{new} is incorrect.  This is due to {\em faults}. A fault is an element $b$ of $R_i$ with $i>1$ such that $y_b = x_{a_1}$.  This is the only element that can suddenly uncover itself when adding $a_{i-1}$ to $R_{i-1}$ on stage II.$(i-1)$, because we have assumed $x$ contains a unique $k$-tuple of equal elements, hence, if $R\in V_1$ is consistent with $x$, no $b$ in $\bigcup R$ satisfies $x_b=x_{a_1}$.%
\footnote{
In fact, this is not a problem even without this assumption. We may adjudge that elements in $x$ equal to $x_{a_1}$ are represented by $\star$ in the assignments.  This justifies \reffoot{overkill}.
}

But since $y$ is a negative input, there are at most $k-1=O(1)$ faults for every choice of $x$. Thus, all we need is to develop a fault-tolerant version of the learning graph from~\reftbl{new} that is capable of dealing with this number of faults.

As an introductory example, consider case $k=3$.  In this case, a fault may only occur in $R_2$.  A fault may come in action only if $y_{a_1}=x_{a_1}$, hence, we may assume there are at most $k-2$ faults in any $y$.  Split $R_2$ into $k-1$ subsets $\{R_2(d)\}_{d\in[k-1]}$.  We know that at least one of them is not faulty, but it is not enough:  we have to assure the contribution from these arcs is $q$ exactly, no matter how many of $R_2(d)$ are faulty, i.e., a variant of~\refeqn{collisionSum}.  We achieve this by splitting $R_1$ into $2^{k-1}-1$ parts $\{R_1(D)\}$ labeled by non-empty subsets $D$ of $[k-1]$.  We uncover an element in $R_2(d)$ if and only if it has a match in $R_1(D)$ for some $D\ni d$.  By adding $a_1$ to $R_1(D)$, we can test whether $\bigcup_{d\in D} R_2(d)$ contains a fault.  This is enough to guarantee~\refeqn{collisionSum} by an application of the inclusion-exclusion principle. The construction in \refsec{construction} is a generalization of this idea for arbitrary $k$.

\subsection{Construction}
\label{sec:construction}
The key vertices of the learning graph are $V_1\cup\cdots \cup V_{k}$, where $V_\stage$ consists of all collections of pairwise disjoint subsets $S = \s{\strut S_i(d_1,d_2,\dots, d_{i-1}, D) }$ labeled by $i\in [k-1]$, $d_j\in [k-j]$, and $\emptyset\subset D\subseteq [k-i]$. There are additional requirements on the sizes of these subsets.

For a non-empty subset $D\subset \N$, let $\mu(D)$ denote the minimal element of $D$. (Actually, any fixed element of $D$ works as well.)  For each sequence $(D_1,\dots,D_{\stage-1})$, where $D_i$ is a non-empty subset of $[k-i]$, let $V_\stage(D_1,\dots,D_{\stage-1})$ consist of all collections $\s{\strut S_i(d_1,d_2,\dots, d_{i-1}, D) }$ such that 
\[
|S_i(d_1,\dots,d_{i-1}, D)| = \begin{cases}
r_i+1,& \mbox{$i<\stage$, $d_1=\mu(D_1),\dots, d_{i-1}=\mu(D_{i-1})$, and $D=D_i$;}\\
r_i,&\mbox{otherwise.}
\end{cases}
\]
Finally, let $V_\stage$ be the union of $V_\stage(D_1,\dots,D_{\stage-1})$ over all choices of $(D_1,\dots,D_{\stage-1})$. 

Again, a vertex in $R = \s{\strut R_i(d_1,d_2,\dots, d_{i-1}, D) }\in V_1$ completely specifies the internal randomness. For each of them, we fix an arbitrary order $t_1,\dots, t_r$ of elements in $\bigcup R$ so that all elements of $R_i$ precede all elements of $R_{i+1}$ for all $i\le k-2$. We say $R$ is consistent with $x$, if $\{a_1,\dots,a_k\}$ is disjoint from $\bigcup R$. Let $q$ be the inverse of the number of $R\in V_1$ consistent with $x$. (Clearly, this number is the same for all choices of $x$.)

The elements still are loaded in the order from~\refeqn{t}. We use a similar convention to name the arcs of the learning graph as in \refsec{first}. Arcs of stages I.$\stage$ are of the form $A^{(R\cap \{t_1,\dots, t_\ell \}, R)}_{t_{\ell+1}}$ for $R\in V_1$ and $0\le \ell < r$. Here, $R\cap T = \s{\strut S_i(d_1,d_2,\dots, d_{i-1}, D) }$ is defined by $S_i(d_1,d_2,\dots, d_{i-1}, D) = R_i(d_1,d_2,\dots, d_{i-1}, D)\cap T$.  Arcs of stage II.$\stage$ are of the form $A^R_j$ with $R\in V_\stage$ and $j\notin\bigcup R$. 

For any $x\in f^{-1}(1)$ and $R\in V_1$ consistent with $x$, the following arcs are taken.  On stage I.$\stage$, for $\stage\in [k-1]$, these are arcs $A^{(R\cap \{t_1,\dots, t_\ell\}, R)}_{t_{\ell+1}}$, where $t_{\ell+1}$ belongs to one of $R_\stage$. On stage II.$\stage$, for $\stage\in[k]$, we have many arcs loading $a_\stage$. For each choice of $(D_i)_{i\in[\stage-1]}$ where $D_i$ is a non-empty subset of $[k-i]$, the arc 
$A_{a_\stage}^{R[D_1\leftarrow a_1,\dots,D_{\stage-1}\leftarrow a_{\stage-1}]}$ is taken where $R[D_1\leftarrow a_1,\dots,D_{\stage-1}\leftarrow a_{\stage-1}] = \s{\strut S_i(d_1,d_2,\dots, d_{i-1}, D) }$ is defined as follows:
\[
S_i(d_1,\dots,d_{i-1}, D) = 
\begin{cases}
R_i(d_1,\dots,d_{i-1}, D)\cup\{a_i\},& \mbox{$i<\stage$, $d_1=\mu(D_1),\dots, d_{i-1}=\mu(D_{i-1})$, and $D=D_i$;}\\
R_i(d_1,\dots,d_{i-1}, D),&\mbox{otherwise.}
\end{cases}
\]

Again, for each arc $A^v_j$, we define a positive semi-definite matrix $X^v_j$ so that $X_j$ in~\refeqn{adversary} are given by $\sum_v X^v_j$. Fix an arc $A^v_j$ and let $S=\s{\strut S_i(d_1,d_2,\dots, d_{i-1}, D) }$ be the set of loaded elements. This time, we define an assignment on $S$ as a function $\alpha\colon \bigcup S\to [m]\cup\{\star\}$ such that $\star\notin \bigcup_D\alpha(S_1(D))$, and, for all $i>1$ and all possible choices of $d_1,\dots,d_{i-1}$ and $D$:
\[\alpha(S_i(d_1,d_2,\dots,d_{i-1},D)) \subseteq \{\star\}\cup \bigcup_{K\ni d_{i-1}} \alpha(S_{i-1}(d_1,\dots,d_{i-2}, K)). \]

\begin{figure}[tb]
\centering \includegraphics[width=9cm]{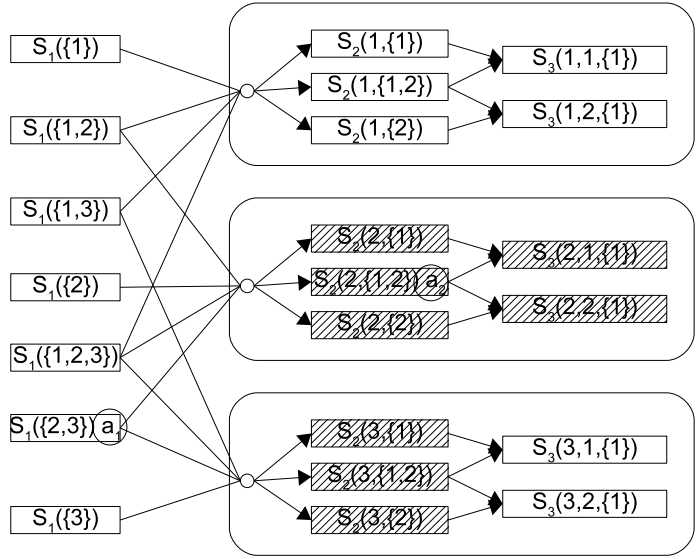} 
\caption{A structure of a vertex of a learning graph for 4-distinctness. The vertex belongs to $V_2(\{2,3\},\{1,2\})$. If there is an arrow between two subsets, a match in the first one is enough to uncover an element in the second one. After $a_1$ is added to $S_1(\{2,3\})$ and $a_2$ is added to $S_2(2,\{1,2\})$, $x$ and $y$ disagree if there is a fault in one of the hatched subsets.}
\label{fig:figa}
\end{figure}

An input $z\in [m]^n$ satisfies assignment $\alpha$ iff, for each $t\in \bigcup S$,
\[\alpha(t) = \begin{cases}
z_t,& \mbox{$t\in S_1(D)$ for some $D$;}\\
z_t,& \mbox{$t\in S_i(d_1,\dots,d_{i-1},D)$ and $z_t\in \bigcup_{K\ni d_{i-1}} \alpha(S_{i-1}(d_1,\dots,d_{i-2},K))$;}\\
\star,& \mbox{otherwise}.
\end{cases}\]
Again, we say inputs $x$ and $y$ agree on $S$, if they satisfy the same assignment $\alpha$.  An example of this construction may be found in \reffig{figa}.

Like before, we define $X^v_j$ as $\sum_\alpha Y_\alpha$ where the sum is over all assignments $\alpha$ on $S$. For the arcs on stage I.1, $Y_\alpha$ are defined as in~\refeqn{tablica1}, and the arcs on stage I.$\stage$, for $\stage>1$, are defined as in~\refeqn{tablica} with $\alpha(S_{\stage-1})$ replaced by $\bigcup_{K\ni d_{\stage-1}} \alpha(S_{\stage-1}(d_1,\dots,d_{\stage-2}, K))$. 

Now consider stage II.$\stage$. Let $A_j^S$ be an arc with $S\in V_\stage(D_1,\dots,D_{\stage-1})$. In this case, $Y_\alpha = q \psi\psi^*$ where
\[
\psi\elem{z} = \begin{cases}
1/{\sqrt{w}},& \text{$f(z)=1$, and $z$ satisfies $\alpha$ and the arc $A_j^S$;}\\
\sqrt{w},& \text{$f(z)=0$, $z$ satisfies $\alpha$, and $\stage+|D_1|+\cdots+|D_{\stage-1}|$ is odd;}\\
-\sqrt{w},& \text{$f(z)=0$, $z$ satisfies $\alpha$, and $\stage+|D_1|+\cdots+|D_{\stage-1}|$ is even;}\\
0,&\text{otherwise.}
\end{cases}
\]
Thus, depending on the parity of $\stage+|D_1|+\cdots+|D_{\stage-1}|$, $X_j^S$ consists of the blocks of one of the following two types:
\begin{equation}
\label{eqn:tablicaFinal}
\begin{array}{r|cc|}
& x & y \\
\hline
x & q/w & q\\
y & q & qw\\
\hline
\end{array}\qquad\mbox{or}\qquad
\begin{array}{r|cc|}
& x & y \\
\hline
x & q/w & -q\\
y & -q & qw\\
\hline
\end{array}
\end{equation}

\paragraph{Complexity} Before we go on proving the correctness of this modified learning graph, let us consider the complexity issue.  The complexity analysis follows the same lines as in \refsec{complexity}. The complexity of stages I.$s$ is proved similarly, by taking $R_i = \bigcup_{d_1,\dots,d_{i-1}, D} R_i(d_1,\dots,d_{i-1},D)$, and noting that $|R_i| = O(k!)r_i = O(r_i)$. Of course, having a match in $R_{i-1}$ is not sufficient for an element in $R_i$ to be uncovered, but this only reduces the complexity. The analysis of stage II.$\stage$ is also similar, but this time instead of one arc loading element $a_\stage$ for a fixed choice of $x$ and $R\in V_1$, there are $2^{O(k^2)}=O(1)$ of them.

\subsection{Feasibility} 
Fix inputs $x\in f^{-1}(1)$ and $y\in f^{-1}(0)$, and let $R\in V_1$ be a choice of the internal randomness consistent with $x$. Compared to the learning graph in \refsec{first}, for a fixed $j\in [n]$, many arcs of the form $A^v_j$ may be taken, thus, we have to modify the $Z_j$ notation. Let $\cZ$ be the set of arcs taken for this choice of $x$ and $R$. The complete list is in \refsec{construction}. We prove that
\begin{equation}
\label{eqn:sum} 
\sum_{A^v_j\in \cZ\colon x_j\ne y_j} X^v_j\elem{x,y} = q.
\end{equation}
Since, again, no arc is taken for two different choices of $R\in V_1$, this proves feasibility~\refeqn{advFeasible}. 

If $x$ and $y$ disagree on $R$ then~\refeqn{sum} holds. The reason is similar to the proof of \refthm{collision}. Again, it is not hard to check that there exists $i\in[r]$ such that $x$ and $y$ disagree on $R\cap \{t_1,\dots,t_{i'}\}$ if and only if $i'\ge i$. Let $j=t_i$, $T=\{t_1,\dots,t_{i-1}\}$, $S = R\cap T$ and $S' = R\cap (T\cup\{j\})$. We claim that $X^{(S,R)}_j\elem{x,y}=q$ and $x_j\ne y_j$. 

Indeed, let $\alpha$ be the assignment $x$ and $y$ both satisfy on $S$, and let $\alpha_x$ and $\alpha_y$ be the assignments $x$ and $y$, respectively, satisfy on $S'$. By the order imposed on the elements in \refeqn{t}, we get that $\alpha(t)=\alpha_x(t)=\alpha_y(t)$ for all $t\in T$. Since $x$ and $y$ disagree on $S'$, it must hold that $\alpha_x(j)\ne \alpha_y(j)$. Hence, $x_j\ne y_j$, and at least one of the is not represented by $\star$ in the assignment on $S'$. Thus, $X^{(S,R)}_j\elem{x,y}=q$ by~\refeqn{tablica1} or~\refeqn{tablica}, in dependence on whether $A^{(S,R)}_j$ belongs to stage I.1 or not.

We claim the contribution to the sum in~\refeqn{sum} from the arcs in $\cZ$ loading $t_{i'}$ for $i'\in [r+k]\setminus\{i\}$ is zero. For $i'>i$, this follows from that $x$ and $y$ disagree before loading $t_{i'}$.  Now consider $i'<i$.  Inputs $x$ and $y$ agree on $S = R\cap\{t_1,\dots,t_{i'}\}$.  Let $j'=t_{i'}$ and $\alpha$ be the assignment $x$ and $y$ both satisfy on $S$.  We have either $x_{j'}=y_{j'}$, or they both are represented by $\star$ in $\alpha$.  In both cases, the contribution is zero (in the second case, by~\refeqn{tablica}).

Now assume $x$ and $y$ agree on $R$. The contribution to~\refeqn{sum} from the arcs of stages I.$\stage$ is 0 by the same argument as in the previous paragraph. Let $\stage$ be the first element such that $x_{a_\stage}\ne y_{a_\stage}$. We claim that if $\stage'\ne\stage$, the contribution to~\refeqn{sum} from the arcs $A^S_{a_{\stage'}}\in \cZ$ with $S\in V_{\stage'}$ is 0.

Indeed, if $s'<s$ then $x_{a_{\stage'}}=y_{a_{\stage'}}$. If $\stage'>\stage$, for each choice of $(D_i)_{i\in[\stage'-1]}$, $x$ and $y$ disagree on $R[D_1\gets a_1,\dots, D_{\stage'-1}\gets a_{\stage'-1}]$, because, by construction, all $a_i$ with $i<\stage'$ are uncovered in the assignment of $x$.

The total contribution from the arcs $A^S_{a_\stage}\in \cZ$ with $S\in V_\stage$ is $q$. This is a special case of \reflem{1} below. Before stating the lemma we have to introduce additional notations. For a vertex $S = R[D_1\leftarrow a_1,\dots,D_\ell\leftarrow a_\ell]$ of the learning graph with $\ell<\stage$, let the {\em block} on this vertex be defined as the set of vertices
\[\cB(S) = \left\{R[D_1\gets a_1,\dots, D_{\stage-1}\leftarrow a_{\stage-1}] \mid \mbox{$\emptyset\subset D_i\subseteq [k-i]$ for $i=\ell+1,\dots,\stage-1$}\right\}. \]
Also, define the {\em contribution} of the block on this vertex as
$\cC(S) = \sum_{S'\in \cB(S)} X^{S'}_{a_\stage}\elem{x,y}$. We prove the following lemma by induction on $\stage-\ell$:
\begin{lem}
\label{lem:1}
Let $R$ and $s$ be as above.  If $x$ and $y$ agree on $S = R[D_1\leftarrow a_1,\dots,D_\ell\leftarrow a_\ell]$ then the contribution from the block on $S$ is $(-1)^{\ell+|D_1|+\cdots+|D_{\ell}|} q$. Otherwise, it is 0.
\end{lem}

Note that if $\ell=0$, the lemma states that the contribution of the block on $R$ is $q$. But this block consists of all arcs of the form $A^S_{a_\stage}$ from $\cZ$. Thus, this proves~\refeqn{sum}.

\pfstart[Proof of \reflem{1}]
If $x$ and $y$ disagree on $S$, they disagree on any vertex from the block, hence, the contribution is 0. 

Now assume $x$ and $y$ agree on $S$. If $\ell=\stage-1$, there is only $S$ in the block. Hence, the contribution is $(-1)^{\ell+|D_1|+\cdots+|D_{\ell}|} q$ by~\refeqn{tablicaFinal}, because $x$ and $y$ agree on $S$ and $x_{a_\stage}\ne y_{a_\stage}$. Now assume $\ell<\stage-1$, and the lemma holds for $\ell$ replaced by $\ell+1$. The block $\cB(S)$ can be expressed as the following disjoint union:
\[\cB(S) = \bigsqcup_{\emptyset\subset D_{\ell+1}\subseteq [k-\ell-1]} \cB(R[D_1\leftarrow a_1,\dots,D_\ell\leftarrow a_\ell,D_{\ell+1}\leftarrow a_{\ell+1}]).\]

Let $I$ be the set of $i\in [k-\ell-1]$ such that $\bigcup_{D} R_{\ell+2}(\mu(D_1),\dots,\mu(D_\ell),i, D)$ does not contain a fault. It is not hard to see that $x$ and $y$ agree on $R[D_1\leftarrow a_1,\dots,D_{\ell+1}\leftarrow a_{\ell+1}]$ if and only if $D_{\ell+1}\subseteq I$. Since $y_{a_1}=\cdots=y_{a_{\stage-1}}=x_{a_1}$ and there is at most $k-1$ element in $y$ equal to $x_{a_1}$, there are at most $k-1-(\stage-1) < k-\ell-1$ faults. Hence, $I$ is non-empty. Using the inductive assumption,
\begin{align*}
\cC(S) &= \sum_{\emptyset\subset D_{\ell+1}\subseteq [k-\ell-1]} \cC(R[D_1\leftarrow a_1,\dots,D_\ell\leftarrow a_\ell,D_{\ell+1}\leftarrow a_{\ell+1}]) \\
&= \sum_{\emptyset\subset D_{\ell+1}\subseteq I} (-1)^{\ell+1+|D_1|+\cdots+|D_{\ell+1}|}q = (-1)^{\ell+|D_1|+\cdots+|D_{\ell}|}q,
\end{align*}
by inclusion-exclusion.
\pfend

\section{Conclusion}
A quantum query algorithm for $k$-distinctness is presented in the paper.  The algorithm uses the learning graph framework.  The improvement in complexity is due to a sequence of new ideas enhancing the framework:  partial assignments in the vertices of the learning graph,  arcs with the weight dependent on the variable being loaded,  fault-tolerant learning graphs, and others.

The future research may concentrate on the following problems.  Is it possible to use some of these ideas to improve the quantum query complexity of other problems?  The complexity of the algorithm in the paper has rather bad dependence on $k$.  Is it possible to improve the dependence using a more advanced fault-tolerance technique?  Finally, we know that the Ambainis' algorithm can be implemented time-efficiently.  Is this true for the algorithm in this paper?

\section*{Acknowledgments}
I am grateful to Robin Kothari for sharing his construction of learning graphs with different arc weights for different values of the variable being loaded and to Andris Ambainis for sharing his algorithm for the graph collision problem that has mostly triggered this research. Also, I would like to thank Andris Ambainis and the anonymous referees for the significant help in improving the presentation of the paper. 

This work has been supported by the European Social Fund within the project ``Support for Doctoral Studies at University of Latvia'' and by FET-Open project QCS.

\bibliographystyle{../../habbrv}
\bibliography{../../bib}

\begin{thebibliography}{10}

\bibitem{shi:collisionLower}
S.~Aaronson and Y.~Shi.
\newblock Quantum lower bounds for the collision and the element distinctness
  problems.
\newblock {\em Journal of the ACM}, 51(4):595--605, 2004.

\bibitem{ambainis:adversary}
A.~Ambainis.
\newblock Quantum lower bounds by quantum arguments.
\newblock {\em Journal of Computer and System Sciences}, 64(4):750--767, 2002,
  \href{http://xxx.lanl.gov/abs/quant-ph/0002066}{{\ttfamily
  arXiv:quant-ph/0002066}}.

\bibitem{ambainis:collisionLower}
A.~Ambainis.
\newblock Polynomial degree and lower bounds in quantum complexity: Collision
  and element distinctness with small range.
\newblock {\em Theory of Computing}, 1:37--46, 2005,
  \href{http://xxx.lanl.gov/abs/quant-ph/0305179}{{\ttfamily
  arXiv:quant-ph/0305179}}.

\bibitem{ambainis:distinctness}
A.~Ambainis.
\newblock Quantum walk algorithm for element distinctness.
\newblock {\em SIAM Journal on Computing}, 37:210--239, 2007,
  \href{http://xxx.lanl.gov/abs/quant-ph/0311001}{{\ttfamily
  arXiv:quant-ph/0311001}}.

\bibitem{ambainis:personal}
A.~Ambainis.
\newblock Personal communication, 2011.

\bibitem{belovs:learning}
A.~Belovs.
\newblock Span programs for functions with constant-sized 1-certificates.
\newblock In {\em Proc. of 44th ACM STOC}, pages 77--84, 2012,
  \href{http://xxx.lanl.gov/abs/1105.4024}{{\ttfamily arXiv:1105.4024}}.

\bibitem{lee:learningKdistPrior}
A.~Belovs and T.~Lee.
\newblock Quantum algorithm for $k$-distinctness with prior knowledge on the
  input.
\newblock 2011,  \href{http://xxx.lanl.gov/abs/1108.3022}{{\ttfamily
  arXiv:1108.3022}}.

\bibitem{belovs:learningClaws}
A.~Belovs and B.~Reichardt.
\newblock Span programs and quantum algorithms for $st$-connectivity and claw
  detection.
\newblock 2012,  \href{http://xxx.lanl.gov/abs/1203.2603}{{\ttfamily
  arXiv:1203.2603}}.

\bibitem{spalek:kSumLower}
A.~Belovs and R.~{\v Spalek}.
\newblock Adversary lower bound for the $k$-sum problem.
\newblock 2012,  \href{http://xxx.lanl.gov/abs/1206.6528}{{\ttfamily
  arXiv:1206.6528}}.

\bibitem{boyer:groverTight}
M.~Boyer, G.~Brassard, P.~{H\o yer}, and A.~Tapp.
\newblock Tight bounds on quantum searching.
\newblock {\em Fortschritte der Physik}, 46(4-5):493--505, 1998,
  \href{http://xxx.lanl.gov/abs/quant-ph/9605034}{{\ttfamily
  arXiv:quant-ph/9605034}}.

\bibitem{brassard:counting}
G.~Brassard, P.~H{\o}yer, and A.~Tapp.
\newblock Quantum counting.
\newblock In {\em Proc. of 25th ICALP}, pages 820--831. Springer, 1998,
  \href{http://xxx.lanl.gov/abs/quant-ph/9805082}{{\ttfamily
  arXiv:quant-ph/9805082}}.

\bibitem{buhrman:querySurvey}
H.~Buhrman and R.~de~Wolf.
\newblock Complexity measures and decision tree complexity: a survey.
\newblock {\em Theoretical Computer Science}, 288:21--43, 2002.

\bibitem{buhrman:productVerification}
H.~Buhrman and R.~{\v{S}}palek.
\newblock Quantum verification of matrix products.
\newblock In {\em Proc. of 17th ACM-SIAM SODA}, pages 880--889, 2006,
  \href{http://xxx.lanl.gov/abs/quant-ph/0409035}{{\ttfamily
  arXiv:quant-ph/0409035}}.

\bibitem{childs:subsetFinding}
A.~Childs and J.~Eisenberg.
\newblock Quantum algorithms for subset finding.
\newblock {\em Quantum Information \& Computation}, 5(7):593--604, 2005,
  \href{http://xxx.lanl.gov/abs/quant-ph/0311038}{{\ttfamily
  arXiv:quant-ph/0311038}}.

\bibitem{dorn:associativity}
S.~D{\"o}rn and T.~Thierauf.
\newblock The quantum query complexity of algebraic properties.
\newblock In {\em Proc. of 16th FCT}, volume 4639, pages 250--260.
  Springer-Verlag, 2007,  \href{http://xxx.lanl.gov/abs/0705.1446}{{\ttfamily
  arXiv:0705.1446}}.

\bibitem{hoyer:adversaryNegative}
P.~H{\o}yer, T.~Lee, and R.~{\v S}palek.
\newblock Negative weights make adversaries stronger.
\newblock In {\em Proc. of 39th ACM STOC}, pages 526--535, 2007,
  \href{http://xxx.lanl.gov/abs/quant-ph/0611054}{{\ttfamily
  arXiv:quant-ph/0611054}}.

\bibitem{jeffery:matrixMultiplication}
S.~Jeffery, R.~Kothari, and F.~Magniez.
\newblock Improving quantum query complexity of boolean matrix multiplication
  using graph collision.
\newblock In {\em Proc. of 39th ICALP}, pages 522--532. Springer, 2012,
  \href{http://xxx.lanl.gov/abs/1112.5855}{{\ttfamily arXiv:1112.5855}}.

\bibitem{kothari:personal}
R.~Kothari.
\newblock Personal communication, 2011.

\bibitem{kutin:collisionLower}
S.~Kutin.
\newblock Quantum lower bound for the collision problem with small range.
\newblock {\em Theory of Computing}, 1(1):29--36, 2005.

\bibitem{lee:learningSubgraphs}
T.~Lee, F.~Magniez, and M.~Santha.
\newblock A learning graph based quantum query algorithm for finding
  constant-size subgraphs.
\newblock 2011,  \href{http://xxx.lanl.gov/abs/1109.5135}{{\ttfamily
  arXiv:1109.5135}}.

\bibitem{lee:stateConversion}
T.~Lee, R.~Mittal, B.~Reichardt, R.~{\v{S}}palek, and M.~Szegedy.
\newblock Quantum query complexity of the state conversion problem.
\newblock In {\em Proc. of 52nd IEEE FOCS}, pages 344--353, 2011,
  \href{http://xxx.lanl.gov/abs/1011.3020}{{\ttfamily arXiv:1011.3020}}.

\bibitem{magniez:walkSearch}
F.~Magniez, A.~Nayak, J.~Roland, and M.~Santha.
\newblock Search via quantum walk.
\newblock In {\em Proc. of 39th ACM STOC}, pages 575--584, 2007,
  \href{http://xxx.lanl.gov/abs/quant-ph/0608026}{{\ttfamily
  arXiv:quant-ph/0608026}}.

\bibitem{magniez:triangle}
F.~Magniez, M.~Santha, and M.~Szegedy.
\newblock Quantum algorithms for the triangle problem.
\newblock {\em SIAM Journal on Computing}, 37(2):413--424, 2007,
  \href{http://xxx.lanl.gov/abs/quant-ph/0310134}{{\ttfamily
  arXiv:quant-ph/0310134}}.

\bibitem{reichardt:adversaryTight}
B.~Reichardt.
\newblock Reflections for quantum query algorithms.
\newblock In {\em Proc. of 22nd ACM-SIAM SODA}, pages 560--569, 2011,
  \href{http://xxx.lanl.gov/abs/1005.1601}{{\ttfamily arXiv:1005.1601}}.

\bibitem{szegedy:walk}
M.~Szegedy.
\newblock Quantum speed-up of markov chain based algorithms.
\newblock In {\em Proc. of 45th IEEE FOCS}, pages 32--41, 2004.

\bibitem{zhu:learning}
Y.~Zhu.
\newblock Quantum query complexity of subgraph containment with constant-sized
  certificates.
\newblock 2011,  \href{http://xxx.lanl.gov/abs/1109.4165}{{\ttfamily
  arXiv:1109.4165}}.

\end{thebibliography}

\end{document}